\PassOptionsToPackage{usenames,dvipsnames}{xcolor}
\documentclass[sigplan,10pt,nonacm]{acmart}
\AtBeginDocument{%
  }

\setcopyright{acmlicensed}
\copyrightyear{2018}
\acmYear{2018}
\acmDOI{XXXXXXX.XXXXXXX}
\acmConference[Conference acronym 'XX]{Make sure to enter the correct
  conference title from your rights confirmation email}{June 03--05,
  2018}{Woodstock, NY}
\acmISBN{978-1-4503-XXXX-X/2018/06}

\newcommand{\IRa}{{SIR}}
\newcommand{\IRb}{{VIR}}

\usepackage[normalem]{ulem}
\usepackage{soul}
\usepackage{tikz}
\usepackage{amsmath}
\usepackage{mdframed}

\usepackage[linesnumbered,ruled,vlined]{algorithm2e}

\usepackage{filecontents}
\usepackage{color}
\usepackage{setspace}
\usepackage[compact]{titlesec}
\usepackage{xspace}
\usepackage{enumitem}
\usepackage{mfirstuc}
\usepackage{caption}
\usepackage[labelformat=simple]{subcaption}

\usepackage{circledsteps}

\usepackage{booktabs}
\usepackage{multirow}
\usepackage{pifont}
\usepackage{tablefootnote}
\usepackage{makecell}
\usepackage{bookmark}
\hypersetup{
  colorlinks=true,
  linkcolor=blue!90!red,
  citecolor=magenta
}
\usepackage{cleveref}
\usepackage[abs]{overpic}
\crefname{section}{§}{§§}

\setenumerate[1]{itemsep=0pt,partopsep=0pt,parsep=\parskip,topsep=5pt}
\setitemize[1]{itemsep=0pt,partopsep=0pt,parsep=\parskip,topsep=5pt}

\begin{document}

\title[]{Retrofitting Control Flow Graphs in LLVM IR for Auto Vectorization}


\author{Shihan Fang}
\affiliation{%
  \institution{Shanghai Jiao Tong University}
  \city{Shanghai}
  \country{China}}
\email{fang-account@sjtu.edu.cn}

\author{Wenxin Zheng}
\affiliation{%
  \institution{Shanghai Jiao Tong University}
  \city{Shanghai}
  \country{China}}
\email{wxzheng98@gmail.com}

\renewcommand{\shortauthors}{}
\newcommand{\ZWX}[1]{\textcolor{blue}{#1}}
\begin{abstract}
Modern processors increasingly rely on SIMD instruction sets, such as AVX and RVV, to significantly enhance parallelism and computational performance. 
However, production-ready compilers like LLVM and GCC often fail to fully exploit available vectorization opportunities due to disjoint vectorization passes and limited extensibility. 
Although recent attempts in heuristics and intermediate representation (IR) designs have attempted to address these problems, efficiently simplifying control flow analysis and accurately identifying vectorization opportunities remain challenging tasks.

To address these issues, we introduce a novel vectorization pipeline featuring two specialized IR extensions: {\IRa}, which encodes high-level structural information, and {\IRb}, which explicitly represents instruction dependencies through data dependency analysis. 
Leveraging the detailed dependency information provided by {\IRb}, we develop a flexible and extensible vectorization framework. 
This approach substantially improves interoperability across vectorization passes and expands the search space for identifying isomorphic instructions, ultimately enhancing both the scope and efficiency of automatic vectorization. 
Experimental evaluations demonstrate that our proposed vectorization pipeline achieves significant performance improvements, delivering speedups of up to 53\% and 58\% compared to LLVM and GCC, respectively.
\end{abstract}

\maketitle

\section{Introduction}

Modern processors use SIMD to execute a single instruction on multiple data elements, enhancing parallelism in fields like scientific computing~\cite{fischer1995practicality}, multimedia processing~\cite{multimedia1, multimedia2}, and machine learning~\cite{limonova2015improving}. 
{
However, using these SIMD units can be challenging for developers. They often need to manually call specific APIs~(Application Programming Interfaces) or ABIs~(Application Binary Interfaces). 
This means they have to write code that directly interacts with these special and hardware interfaces to utilize the SIMD capabilities. 
Handling these details themselves can make the development process more complex and hard to debug.
Mainstream compilers, such as LLVM and GCC, rely on Superword-Level Parallelism (SLP) and Loop-Level Parallelism (LLP) vectorizers to handle independent, isomorphic instructions in straight-line code and loop instructions. 
Unfortunately, these compilers have limited capabilities and exhibit limited extensibility, resulting in missed vectorization opportunities~\cite{llvm-bug-31572, llvm-bug-30787}.
}

Recent research has explored techniques for automatic vectorization.
They mainly focus on making full use of advanced vector-accelerated hardware~\cite{DBLP:conf/pldi/BaghsorkhiVW16,chen2021vegen}, exploring more vectorization opportunities with heuristic~\cite{rocha2020vectorization,10.1145/3631709}, deep reinforcement learning~\cite{10.1145/3368826.3377928, grubisic2023looptune}, machine learning~\cite{stock2012using,ashouri2018survey,springer2018design,tavarageri2021ai} or large language models~\cite{10.1145/3696443.3708929}. These approaches aim to overcome limitations of traditional compilers by identifying more diverse and complex vectorization patterns and adapting to evolving hardware features.



However, these approaches still face significant challenges that limit their effectiveness. 
They can only vectorize a narrow range of code, restricting their overall applicability, and they do not integrate well with optimizations originally designed for single instructions, such as scalar optimizations. 
Additionally, current bottom-up vectorization methods lack flexibility, making it difficult to identify broader similarities between instructions and hindering their ability to detect and exploit instruction-level isomorphism. 
We observe that the key to effective vectorization lies in identifying independent instructions within programs that share structural similarity, a property known as isomorphism. 
However, relying on control-flow graphs (CFG) derived from the intermediate representation (IR) of the code is problematic, as this IR complicates the analysis of instruction dependencies related to control flow. 
Consequently, this approach reduces the efficiency and effectiveness of vectorization.


Traditionally, vectorization is implemented as a pass on LLVM IR, a control-flow-graph-based (CFG-based) representation. To vectorize scalar instructions, we need to move these instructions together, which requires control flow graph (CFG) reconstruction if the instructions are from different basic blocks. However, the reconstructed CFG can be extremely different from the original one, and automatically perform CFG reconstructed is complicated on LLVM IR.
As a result, vectorization is limited to simple code regions, such as within a basic block. 
To enable vectorization across more complex control flows, Predicated Static single-assignment (PSSA)~\cite{chen2022all} is introduced to transform the IR into a non-CFG-based form, replacing the CFG with flat list of instructions and loops and attached \textit{control predicates}~\cite{chen2022all}. Code motion on the flat list is much easier, since we are no longer required to reconstruct the graph to perform code motion, 
Instead of reconstructing the control flow graph for code motion, PSSA~\cite{chen2022all} allows instructions to be moved together directly. This is achieved using a strategy similar to the one commonly used for moving instructions within a basic block, which is much easier.

Nevertheless, constructing PSSA~\cite{chen2022all} directly from LLVM IR is nontrivial. 
The construction still requires complex dominance analysis~\cite{10.1145/390013.808479} to identify loops and CFG reconstruction to convert loops into a \textit{canonical form}~\cite{chen2022all}. And constructing appropriate \textit{control predicates}~\cite{chen2022all} also requires dominance analysis~\cite{10.1145/390013.808479} and even \textit{incomplete heuristics}~\cite{chen2022all}.
In addition, identifying vectorizable instructions is still an open challenge. Inappropriate instruction selection often results in missed opportunities for vectorization.
To address these problems, we introduce two-level IRs and propose a vectorization pipeline from source code or high-level IR.
{
The first level, called {\IRa}, is based on Control Flow Graphs (CFGs) but includes additional information about the high-level structure of the code. The second level, called {\IRb}, is not based on CFGs. Instead, it represents control flow with execution condition (predicates), representing instruction dependencies uniformly with data dependencies, and strengthens loop iteration patterns. 
By using these two levels of IR, it becomes easier to analyze and optimize code for vectorization, which can improve performance on modern processors.}

The level one IR, {\IRa}, is a CFG-based IR preserving high-level structural information.
The difficulty of constructing PSSA~\cite{chen2022all} directly from LLVM IR are mainly about structure analysis, like dominance analysis, and transformation. These steps require recovering structural elements like loop and conditional branch constructs from the CFG in LLVM IR. In contrast, when lowering directly from source code or from a higher-level IR, this structural information is often explicitly retained in the representation and can be directly extracted.
{\IRa} captures loop structure with extracted loop iteration pattern and updating pattern for loop inductive variables. Additionally, loops are converted into a canonical form at this level, which only require simple replacement or insertion.
For branches, {\IRa} captures the branch condition. 
As a result, when constructing non-CFG-based IR from {\IRa}, we can directly utilize these structural information instead of recovering them with complex analysis. What's more, information like loop iteration pattern can be forwarded to non-CFG-based IR, enabling more vectorization opportunities.

The level two IR, {\IRb}, is a non-CFG-based IR derived from PSSA~\cite{chen2022all}. 
On PSSA~\cite{chen2022all}, some optimization algorithms are proposed to perform vectorization across complex control flow such as loop fusion. However, the approach to identify vectorizable instruction remains unclear.
We design a vectorization framework based on {\IRb} to identify vectorization opportunities, estimate the profit, and vectorize instructions.
In addition to \textit{Control Predicates}, we attach an \textit{Iterator} to each instruction and loop.
The \textit{Iterator} we add enable us to identify cross loop vectorization opportunities like loop fusion at instruction level.
We notice that on {\IRb}, control dependence are transformed into data dependence, enabling us to represent the dependence relationship between instructions uniformly. On the one hand, the unified representation allows easier dependence check for instructions. On the other hand, the dependence relationship, always in the form of producer-consumer relationship between instructions provide us with a way to identify chains of vectorizable instructions.  
Inspired by these, our vectorization framework on {\IRb} utilize a data structure we propose to capture the dependence. The framework is both comprehensive, covering the full vectorization workflow on {\IRb}, and extensible, supporting extensions of new analyses and transformation techniques.

We evaluated the effectiveness of {\IRa} using both vectorizable code and general programs. In terms of compilation time, {\IRa} does not introduce additional overhead. When compiling user-specific code, {\IRa} achieves up to a 15\% improvement compared to LLVM, with no more than a 5\% overhead compared to GCC.  For real-world image pixel processing code that existing compilers cannot vectorize, {\IRa} achieves up to a 53\% performance improvement compared to LLVM and up to 58\% compared to GCC.

In summary, the main contributions of this paper are as follows:
\begin{itemize}
    \item We present {\IRa}, an IR derived from Predicated SSA, designed to fully replace control flow with data flow while enhancing the representation of control information.
    \item We propose a flexible vectorization framework that significantly expands the search space for vectorizable instructions.
    \item We demonstrate that our implementation of a vectorizing compiler for simplified C code is comparable to, and often surpasses, the vectorization capabilities of LLVM and GCC, including both their loop and SLP vectorizers.
\end{itemize}
\section{Background and Motivation}

\subsection{Computation under SIMD}

As data processing demands have increased, achieving high performance increasingly depends on exploiting parallelism. 
One effective form of parallelism is data-level parallelism (DLP), where the same operation is applied to multiple data elements simultaneously. 
To support this, modern processors have introduced Single Instruction Multiple Data (SIMD) operations.
SIMD vectorization has shown substantial performance improvements in fields like scientific computing~\cite{fischer1995practicality}, multimedia processing~\cite{multimedia1, multimedia2}, and machine learning~\cite{limonova2015improving}. This process, commonly known as vectorization, has become a standard technique for harnessing SIMD hardware to accelerate data-intensive tasks.

To utilize SIMD, developers traditionally relied on manual assembly coding, intrinsics\cite{riscv-vector-intrinsic}, or specialized libraries~\cite{cppreference-simd}. While manual SIMD code can deliver highly optimized performance, it often demands significant engineering effort and deep understanding of low-level hardware details to achieve effective vectorization.
As the diversity and complexity of compute kernels continue to grow, manual vectorization proves difficult to scale and maintain, motivating the need for automatic techniques.

\subsection{Automatic Vectorization Techniques}

Vectorizing compilers address this by performing optimizations on intermediate representation (IR)~\cite{alfred2007compilers, DBLP:conf/cgo/LattnerA04}. The IR is an abstract and lower-level form of the source code, designed to simplify analysis and transformation. These optimizations are usually applied to control flow graph (CFG)-based IR~\cite{alfred2007compilers,muchnick1997advanced}, where the program is modeled as a graph of basic blocks connected by edges on the control flow. Each basic block consists of a sequence of instructions with a single entry and exit point. The control flow edges represent possible execution paths between blocks and capture the program’s branching structure, including loops, conditionals, and jumps.

Automatic vectorization primarily works at two levels: Loop-Level Parallelism (LLP)~\cite{DBLP:journals/toplas/AllenK87} and Superword-Level Parallelism (SLP)~\cite{larsen2000exploiting}. LLP focuses on vectorizing loops with regular iteration and linear memory access patterns. SLP, on the other hand, detects vectorization opportunities within basic blocks. SLP analyzes instruction dependencies and applies simple code motion to enable vectorization. 

SLP is considered a simpler and more flexible approach to perform vectorization~\cite{chen2022all}, as it does not require complex dependence analysis across loop iterations and new SIMD instructions can be supported by incorporating additional heuristics~\cite{10.1145/2345156.2254106,mendis2018goslp,porpodas2018vw, porpodas2015pslp}.
Additionally, techniques like loop unrolling~\cite{larsen2000exploiting, rocha2020vectorization} complement SLP by exposing additional parallelism across loop iterations, enabling more effective vectorization. With the help of loop unrolling with appropriate unrolling factor, SLP can achieve vectorization performance comparable to LLP.

\begin{figure}[htbp]
    \centering
    \includegraphics[width=\columnwidth]{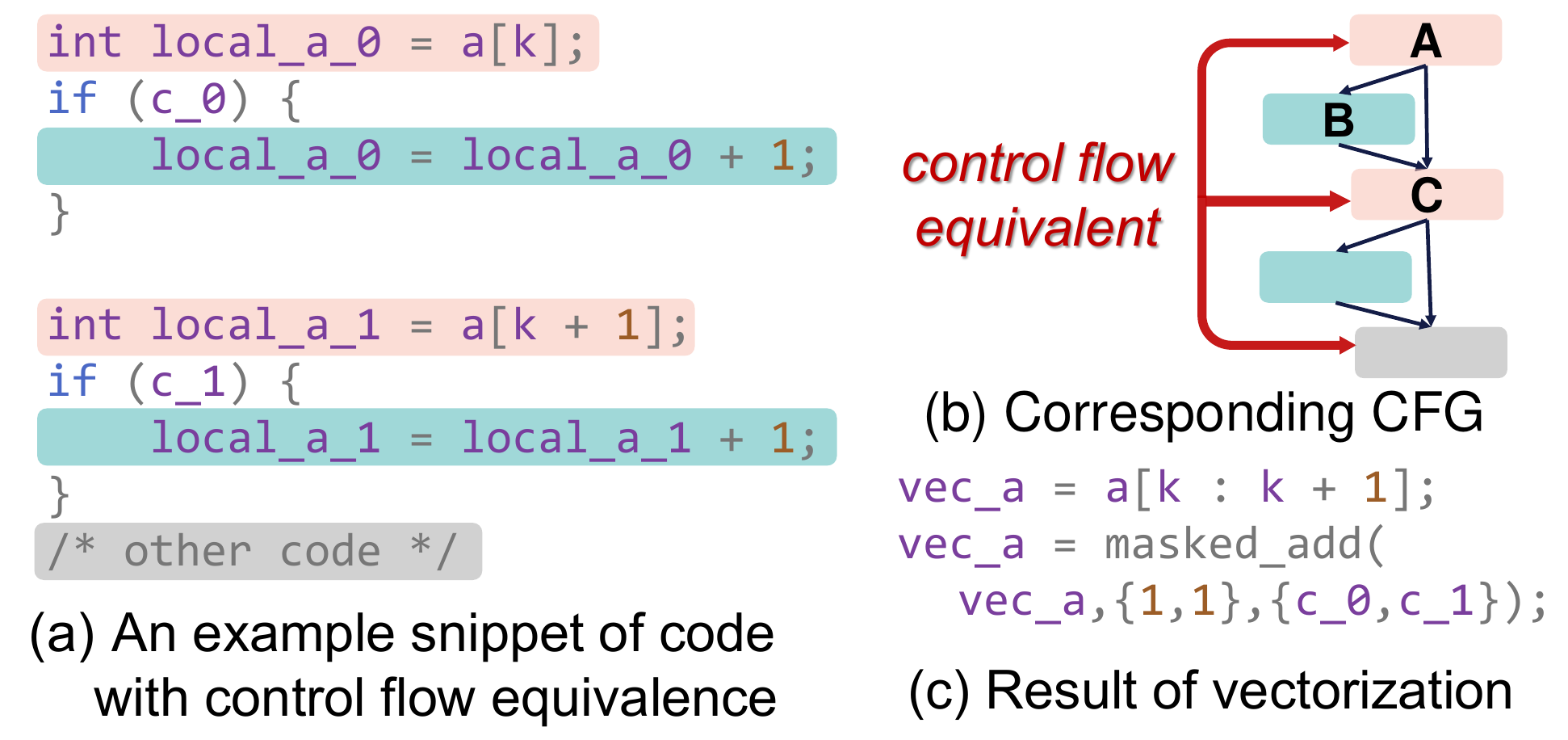}
    \caption{
    An example of code containing control-flow-equivalent~\cite{chen2022all} basic blocks—two blocks execute under identical conditions. The pink basic blocks are control flow equivalent and the two load instructions in these two blocks can be vectorized. The conditional addition on the loaded values can also be vectorized using masked addition, where each element operation executes only if the corresponding mask element is \textit{true}.}
    \label{fig:motiv:control-equivalent}
\end{figure}

However, traditional SLP is limited to basic blocks because vectorizing instructions across basic blocks requires moving them to the same location, which require complex control flow transformation on control-flow-graph-based IR like LLVM IR ~\cite{DBLP:conf/cgo/LattnerA04}.
A notable solution to this problem is Predicated SSA (PSSA), introduced by Chen et al.~\cite{chen2022all}. This approach transforms control flow into a linear sequence of instructions and loops. It replaces control flow with symbolic boolean expressions called \textit{control predicates} attached to each instruction or loop, indicating whether the instruction or loop should execute. PSSA transforms control dependencies on control flow into data dependencies on \textit{control predicates}. Code motions can thus be performed easily on PSSA as long as the dependencies, including data dependencies on operands and \textit{control predicates}, are satisfied. As a result, PSSA is able to exploit vectorization between instructions in different basic blocks and even different loops with advanced techniques like loop fusion and loop co-iteration. Figure\ref{fig:motiv:control-equivalent} gives an example of vectorizing instructions in different basic blocks.

\subsection{Revisiting PSSA in Automatic Vectorization}\label{sec:pssa-limitation}

PSSA is implemented to support vectorization within a pass for LLVM IR~\cite{DBLP:conf/cgo/LattnerA04}. It is first constructed from the original LLVM IR. Optimization analyses and transformations are then performed on the PSSA to enable vectorization. After that, the modified representation is converted back to LLVM IR. Although this approach enables more effective vectorization~\cite{chen2022all}, it still presents several limitations.

\paragraph{Premature lowering to CFG hinders transformation to non-CFG-based IRs like PSSA.} 
The first step to transform LLVM IR into PSSA is detecting the loops and converting them into a canonical form to facilitate the transforming process. Loop detection on LLVM IR requires complicated analysis on CFG ~\cite{muchnick1997advanced} and conversion to the canonical form requires CFG reconstruction including inserting basic blocks as dedicated loop header or pre-header as defined by  Chen et al.~\cite{chen2022all}. Computation for \textit{control predicates} is even more complex. The proposed algorithm utilizes dominance analysis~\cite{10.1145/390013.808479} to infer the \textit{control predicates}. And incomplete heuristic is used to simplify the \textit{control predicates} of blocks with same execution condition, also refered to as control flow equivalent blocks as shown in Figure\ref{fig:motiv:control-equivalent}. 

However, loop structures are trivial in source code and execution conditions can be inferred easily as symbolic boolean expressions, which can be further simplified with fewer effort, from branches in source code.
The complication of transforming to PSSA from LLVM IR comes from the loss of high-level structural information like loop structures when we lower from source code to LLVM IR. Although some structural information can be reconstructed with analysis, these analysis methods always take a lot of effort.

\begin{figure}[htbp]
    \centering
    \includegraphics[width=\columnwidth]{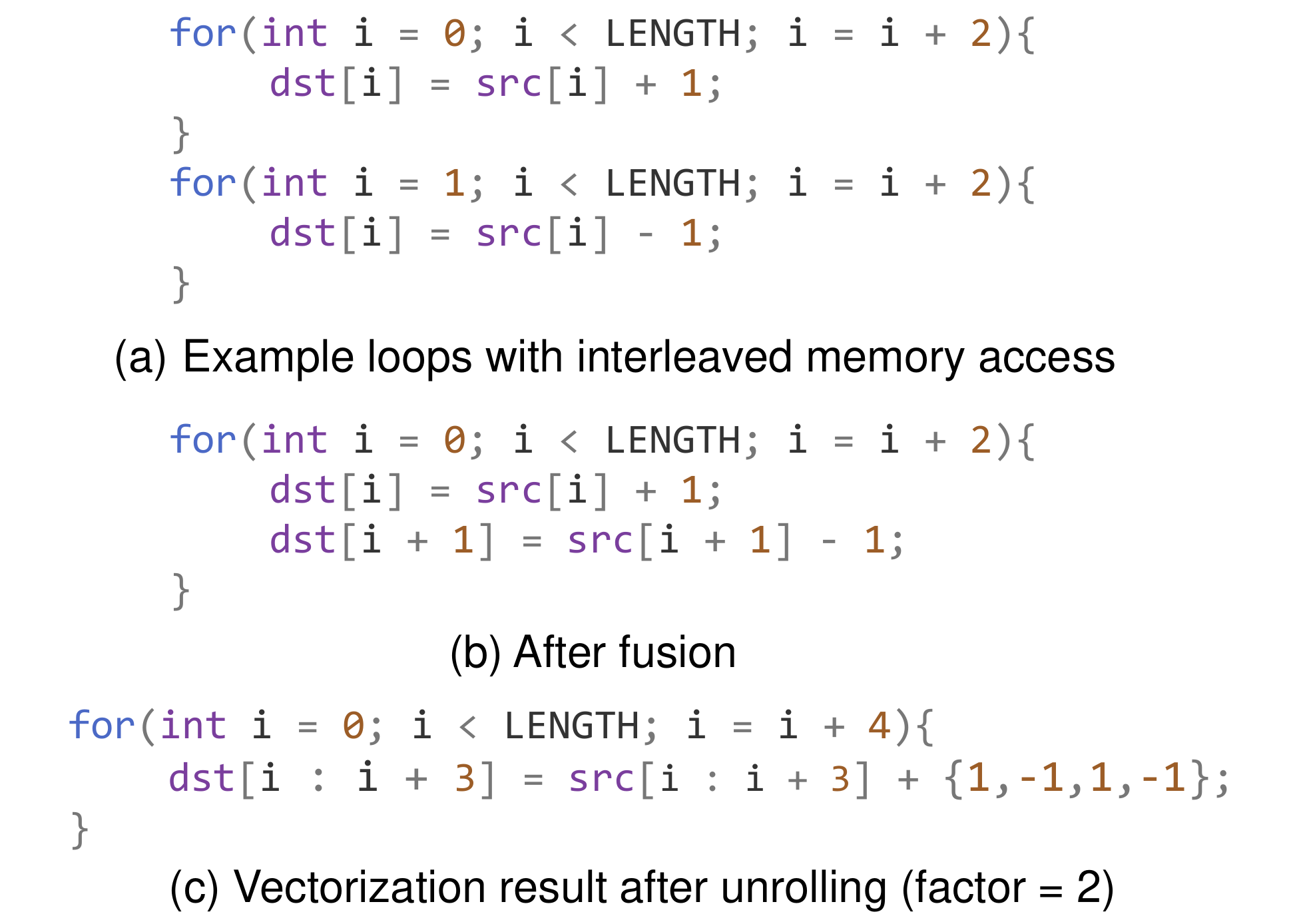}
    \caption{
    An example of code with interleaved access to the same array by processing even and odd indices separately. After applying loop fusion followed by unrolling, the two loops can be fully vectorized.
}
    \label{fig:motiv:fusion-unroll}
\end{figure}

\paragraph{Identifying vectorization opportunities on PSSA is non-trivial.} 
LLP is exploited through loop unrolling, but selecting an appropriate unrolling factor is a challenging task. Chen et al.~\cite{chen2022all} address this by traversing each loop, virtually unrolling it with a set of candidate factors, and estimating the performance benefit to select the optimal one. Since this estimation relies on vectorizing the virtually unrolled code, evaluating all possible factors can be computationally expensive. To reduce the search space, candidate factors are typically limited to those that align with the register width and data type size within the loop body, which helps ensure efficient use of vector resources. However, it considers only individual loops and may lead to suboptimal results when multiple loops interact.
Advanced techniques such as loop fusion and co-iteration have been proposed to uncover additional vectorization opportunities. Yet, identifying effective candidates for these transformations remains an open challenge. Candidates for loop fusion are typically loops with identical iterations, this is also a strong constrain which may lead to suboptimal results. 

The example in Figure\ref{fig:motiv:fusion-unroll} demonstrates these challenges well. Two complementary loops operate on the same arrays, one reads from and writes to the even indices while the other handles the odd indices. Assuming the vector resources can process 4 elements with one instruction. The unrolling algorithm tries to unroll each loop with factor 4. Then, each loop may fail to vectorize or is partially vectorized with \textit{shufflevector} and \textit{extractelement}~\cite{LLVM-LangRef-VectorOps}. Additionally, due to differing iteration patterns, the algorithm may not identify these loops as candidates for fusion. However, after manually fusing these two loops as shown in Figure\ref{fig:motiv:fusion-unroll}(b) and unroll the fused loop with factor 2, we can fully vectorize these loops and make full use of vector resources.

\section{Compilation Pipeline Overview}

\begin{figure}[htbp]
    \centering
    \includegraphics[width=\columnwidth]{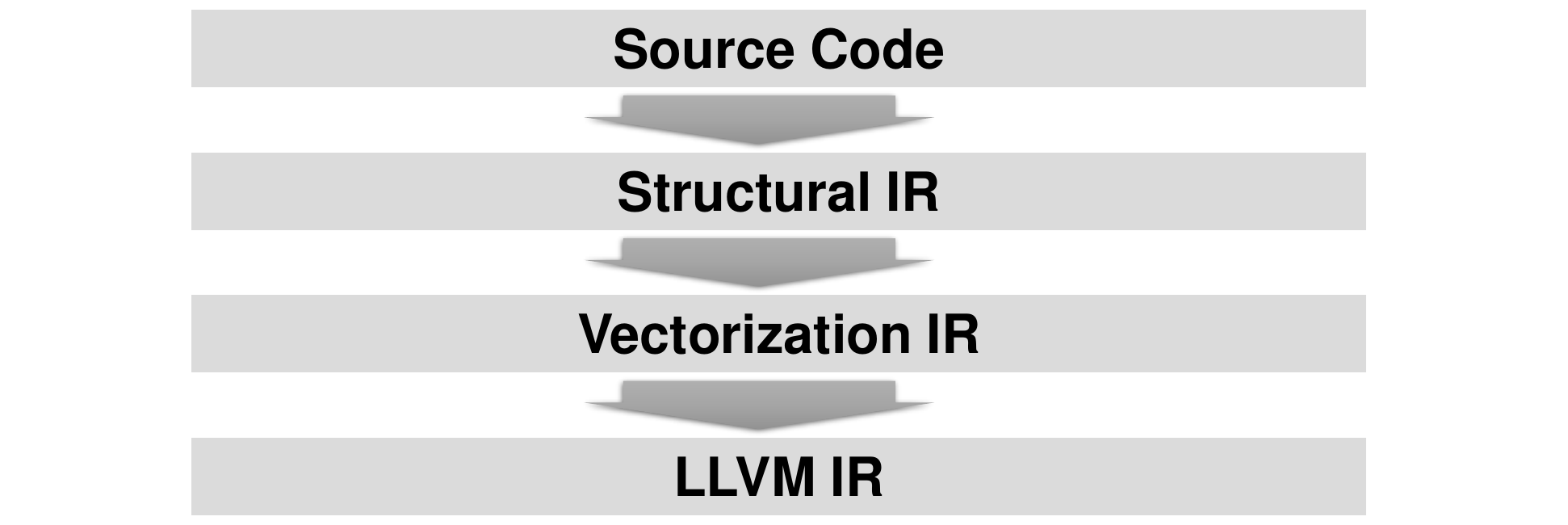}
    \caption{
    Compilation pipeline. 
}
    \label{fig:design:pipeline}
\end{figure}

To address the limitations of premature lowering and efficiently identify vectorization opportunities as discussed in Section~\ref{sec:pssa-limitation}, we introduce a novel compilation pipeline to automatically vectorize source code. 

As demonstrated in Figure~\ref{fig:design:pipeline}, we first transform the source code or high-level IR into {\IRa} (Section~\ref{sec:structural-ir}). {\IRa} is at a higher level than CFG and preserves structural information, including loop structures, to facilitate transformation and optimization. 
Subsequently, we transform {\IRa} to non-CFG-based {\IRb} (Section~\ref{sec:vectorization-ir}) and propose a framework to vectorize on {\IRb} . This framework is designed to be flexible, allowing for easy extensions and more precise identification of vectorization candidates at a finer granularity. After that, {\IRb} is lowered to LLVM IR for further optimization or code generation.

\section{Structural IR ({\IRa})}

Structural IR ({\IRa}) is proposed to preserve high-level information directly collected from source code or high-level IR. 

{\IRa} construction (Section \ref{sec:sir-cnstruction}) from source code or high-level IR is easy to conduct. Additionally, analysis and optimization passes on traditional CFG-based IR can also be performed on {\IRa} with little modification on the algorithms.
After optimizations on {\IRa}, we will show in Section \ref{sec:sir-lowering} that with the help of high-level structural information in {\IRa}, we can construct non-CFG-based {\IRb} more efficiently.

\subsection{Design of {\IRa}}\label{sec:structural-ir}

{\IRa} represents each function in a program with two main parts. One part is a high level structure tree and the other part is a directed graph. 

\begin{figure}[htbp]
    \centering
    \includegraphics[width=1\linewidth]{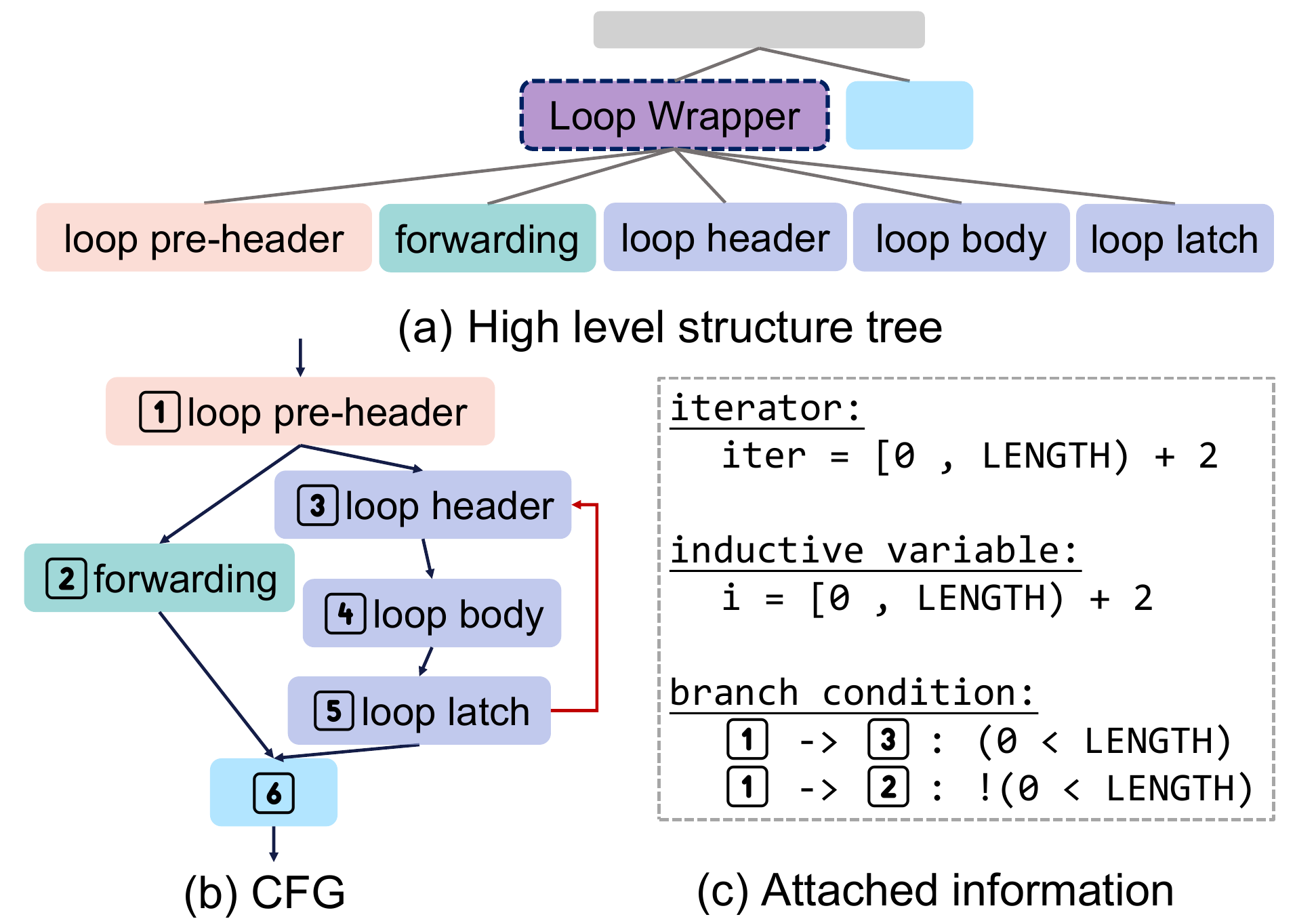}
    \caption{Translated {\IRa} of the first loop in Figure\ref{fig:motiv:fusion-unroll}(a). (a) is a sub-tree representing the loop rooted at a \textit{Loop Wrapper} and a \textit{Block} after the loop. (b) is the corresponding CFG of the \textit{Blocks} in (a). And (c) presents some of the important information we extracted from source code.}
    \label{fig:sir-sample}
\end{figure}

The shared element between the structure tree and a control flow graph is the \textit{Blocks}. These \textit{Blocks} serve as leaves in the structure tree and nodes in the directed graph. They are similar to basic blocks in a traditional control flow graph. In contrast to basic blocks, \textit{Blocks} in {\IRa} do not retain terminal instructions. Instead, the control flow information is recorded on the directed edges of the graph. The edge direction indicates the branch from one block to another. The condition on an edge specifies when a branch should be taken. Since this directed graph maintains control flow information, we also refer to it as a CFG.

The structure tree roots at a \textit{Function Wrapper} and nodes on the tree includes \textit{Loop Wrappers} and \textit{Blocks}. 
Each \textit{Loop Wrapper} is a sub-tree in the structure tree representing a loop in the form of a \textit{do-while} structure from the source code. To facilitate lowering to an IR that is not based on control flow graphs, we further transform each \textit{Loop Wrapper} into a canonical form with:
\begin{itemize}
    \item \textit{loop pre-header}: The precursor of \textit{loop header} typically contains the computation that determines whether the loop should execute at least once. It also performs the initialization of the variables used within the loop.
    \item \textit{loop header}: The unique loop entry and destination block of the back edge in a loop.
    \item \textit{loop latch}: The unique source block of the back edge in a loop.
\end{itemize}

In addition to details about how loops are composed of blocks and nested loops, information such as the loop iteration pattern and inductive variables is attached to the \textit{Loop Wrapper}. This information will be forwarded to non-CFG-based IR where we perform vectorization. 

Figure \ref{fig:sir-sample} demonstrates a {\IRa} example of one of the loops in Figure\ref{fig:motiv:fusion-unroll}. Figure \ref{fig:sir-sample}(a) represents the loop with the \textit{Loop Wrapper} in canonical form. \textit{Blocks} are both tree leaves in the structure tree in Figure \ref{fig:sir-sample}(a) and nodes in the CFG in Figure \ref{fig:sir-sample}(b). Branch conditions on edges are recorded in \ref{fig:sir-sample}(c). Besides that, the iteration pattern and inductive variables in the loop are also shown in \ref{fig:sir-sample}(c). The \textit{iterator} indicates that this is a loop with a regular iteration pattern, making it a perfect candidate for unrolling and possibly vectorization.

\subsection{{\IRa} Construction} \label{sec:sir-cnstruction}

We construct {\IRa} with source code or high-level IR.
As in conventional compilation pipelines, the source code is parsed into an Abstract Syntax Tree (AST)~\cite{alfred2007compilers, muchnick1997advanced}, which explicitly represents structures like branches and loops. 
Other high-level IRs with similar explicit structure information are also compatible with our construction. 

We traverse the AST or high-level IR and transform the structures and build \textit{Blocks} along the way.
Constructing \textit{Loop Wrapper} in canonical form requires transforming the original code structure. But this transformation involves little effort when building from the source code. 
The transformation from a \textit{for} loop or a \textit{while-do} loop into a \textit{do-while} form can be achieved by inserting a conditional check before the loop. This check ensures that if the loop is meant to execute at least once, control jumps to the loop body. Otherwise, it jumps to a {forwarding block}, which is a placeholder with no instructions, as shown in Figure\ref{fig:sir-sample}.

\subsection{Analysis and Optimization on {\IRa}}

We can also conduct analysis or optimization passes on traditional CFG on {\IRa}. Since {\IRa} contains a directed graph similar to traditional CFG, many optimization passes can be applied using algorithms similar to the ones designed for CFGs. 
These passes include constant propagation and dead code elimination~\cite{muchnick1997advanced}, which help simplify the code structure. In addition, {\IRa} can be transformed into Static Single Assignment (SSA)~\cite{cytron1991efficiently} form to facilitate data flow analysis and ease translation to lower-level IR which is necessary for the transformation we will introduce in Section\ref{sec:sir-lowering}.

Additional structural information can be extracted with analysis on {\IRa}. 
Loops iteration patterns can be decided by loop condition and its updating pattern. 
For example, the first \textit{for} loop in Figure\ref{fig:motiv:fusion-unroll}(a) execute when \texttt{i\ \textless\ LENGTH}, since \texttt{i} is initialized with $0$ and increase by $2$ each iteration, the loop iterates between $\left[0, \texttt{LENGTH}\right)
$ with step \texttt{2} as shown in Figure\ref{fig:sir-sample}(c). Loop inductive variables can also be identified through data flow analysis techniques.

It is notable that LLP, enabled by loop unrolling and SLP, is most effective when applied to loops with regular iteration patterns, such as those that traverse a fixed range with a constant stride.
The reason is that these patterns facilitate predictable dependence analysis and efficient work partitioning, both of which are critical to vectorization performance.
Detecting loops with regular iteration patterns efficiently and accurately is therefore essential. 
To identify more such loops, the analysis to extract additional structural information must be extensible.
For example, in some loops, the iterator is updated through a function call rather than explicit arithmetic. 
In these cases, analysis on CFG-based IR alone is often insufficient to determine regularity.
However, through pattern matching or heuristic approaches, it is possible to recognize such loops and expand the set of candidates for LLP vectorization.

\section{Vectorization IR ({\IRb})}

We introduce Vectorization IR ({\IRb}) to improve the expressiveness of Predicated SSA by enabling more precise representation of control flow information. Based on this IR dedicated for vectorization, we introduce a flexible vectorization framework (Section\ref{sec:vectorization-framework}) that identifies vectorization opportunities more precisely and efficiently.

\subsection{Design of \IRb}\label{sec:vectorization-ir}

Similar to Predicated SSA \cite{chen2022all}, {\IRb} is a non-CFG-based IR and replaces the CFG with a flat code list. This list is organized into pairs of \textit{Item} and \textit{Control Predicate}. An \textit{Item} can be either an instruction or a loop, with each loop body introducing a new hierarchical layer in the code list. A \textit{Control Predicate} is a Boolean expression involving variables or constants that determines whether the corresponding \textit{Item} should execute.
To enhance vectorization across complex control flows, particularly for inter-loop instructions, we replace the \textit{Mu Instruction} used for loop induction variables with an \textit{Iota Instruction}, which derives values directly from iteration. Furthermore, each pair of \textit{Control Predicate} and \textit{Item} is extended with an \textit{Iterator} to specify the iteration mode of the parent layer.
The combination of \textit{Control Predicate}, \textit{Iterator}, and \textit{Item} forms an \textit{Entry}, enabling greater flexibility and efficiency in vectorization across complex control flows. If the \textit{Item} is an instruction, the \textit{Entry} is referred to as an \textit{Instruction Entry}. If the \textit{Item} is a loop, it is referred to as a \textit{Loop Entry}.

\begin{figure}[htbp]
    \centering
    \includegraphics[width=1\linewidth]{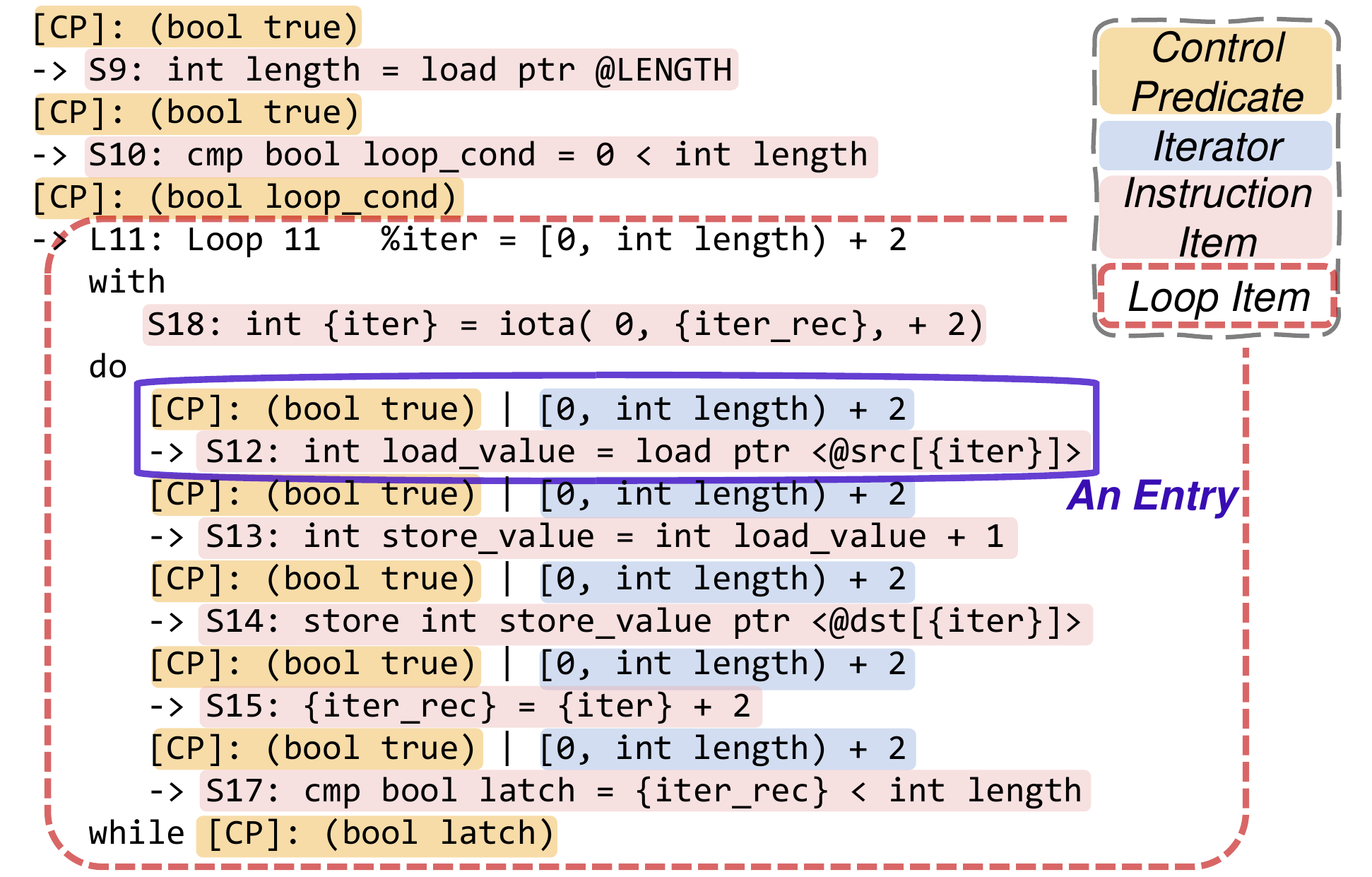}
    \caption{Translated {\IRb} of the first loop in Figure \ref{fig:motiv:fusion-unroll}(a).}
    \label{fig:vir-sample}
\end{figure}

Figure \ref{fig:vir-sample} illustrates a {\IRb} example of one of the loops in Figure \ref{fig:motiv:fusion-unroll}. The comparision instruction \texttt{S10} computes the condition which decides whether the loop \texttt{L11} should execute, so the result of \texttt{S10} compose the \textit{Control Predicate} of the \textit{Loop Entry}. The \textit{Loop Item} is composed of a \texttt{with} list of \textit{Instruction Items} defining variables used in the loop body, a code list serving as the loop body and a \textit{loop latch} which is the \textit{Control Predicate} deciding whether we should exit the loop.
In this example, the \texttt{with} list contains an \textit{Iota Instruction}, indicating that \texttt{\{iter\}} is assigned with \texttt{0} when we execute the loop body for the first time and should be assigned with \texttt{\{iter\_rec\}} for the following iteration. Additionally, the value of \texttt{\{iter\}} is increased by \texttt{2} after each iteration. 
The \textit{Loop Item} introduces a new code list layer, and the \textit{Control Predicates} of \textit{Entries} inside the code lists are irrelevant to the \textit{Control Predicates} of the \textit{Loop Entry}. As shown in Figure \ref{fig:vir-sample}, we assign \texttt{True} to all the entries in the code list, since they will be executed unconditionally in the loop body.

\subsection{{\IRb} Construction from {\IRa}}\label{sec:sir-lowering}

We construct {\IRb} with deep-first traversal on the structure tree in {\IRa}. 
A code list stack is maintained to store the nested layers of code list introduced by nested loops. In addition, since all instructions in a \textit{Block} execute under same condition, which means that the corresponding \textit{Instruction Entries} share the same \textit{Control Predicate}, we maintain a map to record \textit{Control Predicate} shared by \textit{Instruction Entries} from the same \textit{Block}.  
To calculate the \textit{Control Predicate} for each \textit{Entry}, we utilize a \textit{Control Predicate Calculator}. It collects branch conditions during traversal, computes and simplifies the \textit{Control Predicates}

The traversal starts with the \textit{Entry Block} of the function, which is the first block that must execute in the function. 
\textit{Instruction Entries} are constructed from instructions in the \textit{Block} in sequence and will be appended to the code list on top of the code list stack. Since each instruction is not in a loop body and execute under no condition, \texttt{NULL} is assigned to the \textit{Iterator} and \texttt{True} is assigned to the \textit{Control Predicates} of the corresponding \textit{Instruction Entry}. 
After all the instructions in one \textit{Block} are transformed, we switch to next unvisited node following the traversal order. 
If we visit another \textit{Block} $b$, the \textit{Control Predicate Calculator} first find this precursors $Pre$ from the CFG in {\IRa}. For all $p_i\in Pre$, it then transforms the branch condition on the directed edge $p_i \to b$ into \textit{Predicate} $c_i$, which serve as the basic units of \textit{Control Predicates}. Let $cp_i$ be the \textit{Control Predicates} related to \textit{Entries} transformed from $p_i$, the \textit{Control Predicates} of \textit{Entries} transformed from $b$ is calculated as $\bigvee (c_i \land cp_i)$. The \textit{Control Predicate Calculator} can simplify the \textit{Control Predicates}  by performing symbolic computations over boolean expressions. 

Every time we visit a \textit{Loop Wrapper} in the structure tree, we transform the child nodes following the traversal order and construct a \textit{Loop Entry} when we meet \textit{loop header} $b_h$. The \textit{Control Predicate} of the \textit{Loop Entry} is calculated as if we want to get \textit{Control Predicate} of \textit{Entries} transformed from $b_h$. 
If the loop has regular iteration pattern recorded in \textit{Loop Wrapper}, the iterator is assigned to \textit{Iterator} in the \textit{Loop Entry}. A \textit{Loop Entry} introduce a new code list layer, so we push a new code list into the code list stack, which will be popped out after visiting all nodes in the loop, and set \textit{Control Predicates} for \textit{Entries} transformed from $b_h$ to \texttt{True}. 

For example, when transforming from {\IRa} in Figure \ref{fig:sir-sample} to {\IRb} in Figure \ref{fig:vir-sample}, we first visit the \textit{loop pre-header} and transform the instructions. Since both instructions execute unconditionally, the \textit{Control Predicates} are \texttt{True}. The next visited \textit{Block} is \textit{forwarding}. The \textit{Control Predicate Calculator} will record the \textit{Control Predicate} related to it as \texttt{!loop\_cond} because \textit{loop pre-header} branches to \textit{forwarding} when \texttt{loop\_cond} do not holds. 
Then, we visit \textit{loop header} and construct a \textit{Loop Entry} for the loop. The \textit{Control Predicate} related to \textit{loop header} is calculated as \texttt{loop\_cond}, and is assigned to the \textit{Control Predicate} of the \textit{Loop Entry}. 
When traversing the \textit{Blocks} in loop, the base \textit{Control Predicate} is reset to \texttt{True}. As a result, \textit{Control Predicates} of \textit{Instruction Entries} in the loop body are all \texttt{True} as there is no branch in this loop body. 
After traversing the sub-tree rooted at the \textit{Loop Wrapper} in Figure \ref{fig:sir-sample}, we will visit the \textit{Block} after the loop. This \textit{Block} numbered \texttt{6} has  \texttt{2} and \texttt{5} as its precursors, \textit{Control Predicate Calculator} will first calculate the related \textit{Control Predicate} as $($\texttt{!loop\_cond} $\lor$ \texttt{loop\_cond}$)$ and further simplify it as \texttt{True}.
It's notable that instead of using heuristic with dominance analysis to detect control-flow equivalence, our construction allow the control-flow-equivalent \textit{Blocks} \texttt{1} and \texttt{6} to share the same \textit{Control Predicate} by nature.

\subsection{Vectorization Framework}\label{sec:vectorization-framework}

\begin{figure*}[htbp]
    \centering
    \includegraphics[width=1\linewidth]{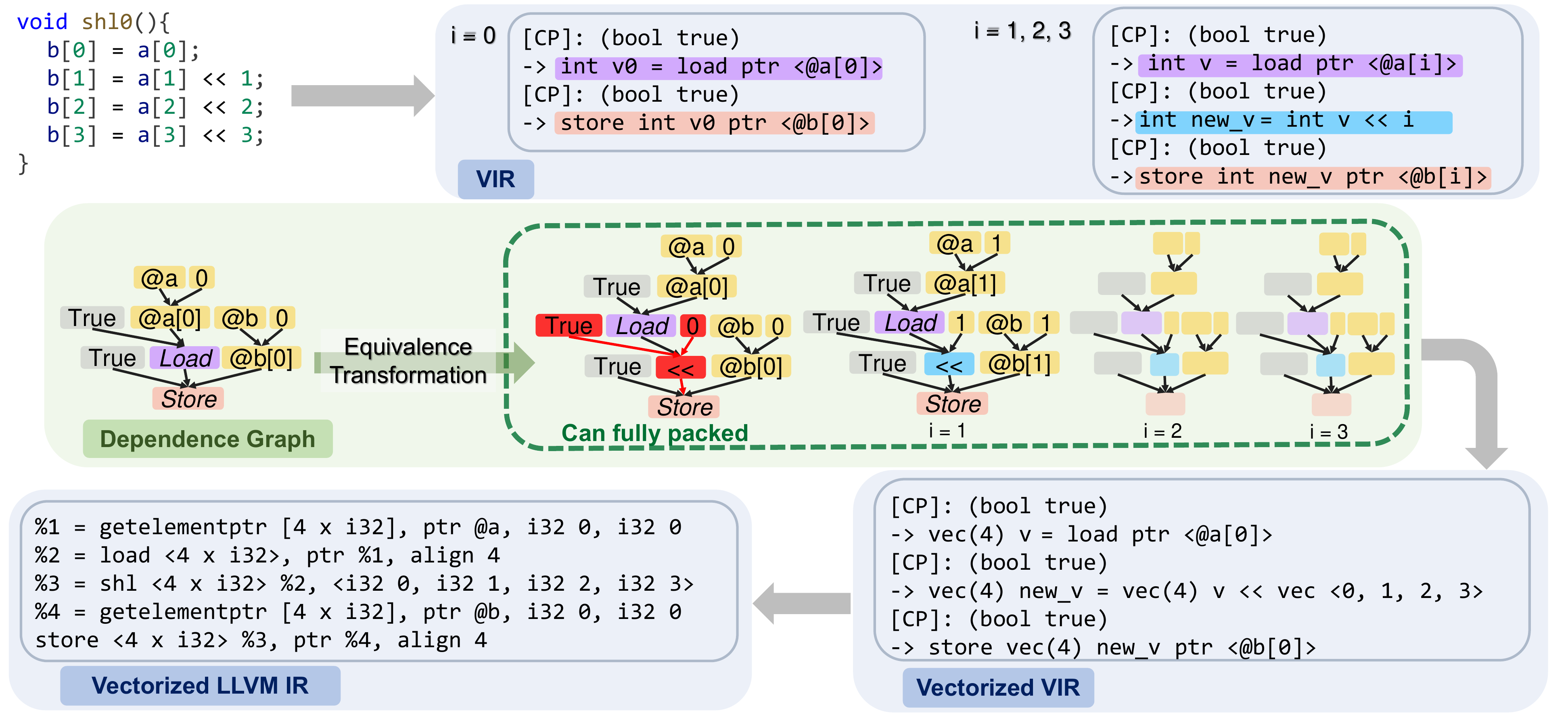}
    \caption{Demonstration of vectorizing scalar code with {\IRb}. Unlike traditional SLP, which vectorizes only a pair of similar operations, our framework on {\IRb} exploits isomorphism through equivalence transformations to fully vectorize all four operations.}
    \label{fig:vectorization}
\end{figure*}

Vectorization with SLP starts with identify independent isomorphic instructions as candidates.
Isomorphic instructions are defined as instructions with the same operations in the same order.~\cite{larsen2000exploiting} 
Instruction dependence can be classified as control dependence and data dependence. With the help of \textit{Control Predicate}, we replace control dependence with data dependence, enabling us to represent dependence on {\IRb} uniformly. 
Unified dependence representation makes checking whether some instructions are independent easier.
Additionally, most of the dependence comes from the producer-consumer relationship between instructions, making this representation ideal for identifying chains of vectorizable instructions and uncovering extended opportunities for vectorization.
Inspired by these, we designed a \textit{dependence graph} to capture the dependence between \textit{Entries} in each layer's code list and utilize it as the central structure in our vectorization framework.

Vectorization in {\IRb} starts with constructing a \textit{dependence graph} (Section\ref{vir:dependence-graph}) to capture relationships among \textit{Entries} within each layer.
Candidates for vectorization are detected on the dependence graph with pattern matching techniques to identify isomorphism and we refer to these candidates as {Instruction Packs} (Section\ref{sec:vector-packing}).
Section\ref{sec:packing-extension} will show that we can flexibly make extensions in our framework to detect more {Instruction Packs}. 
Next, candidate loop unrolling factors for loops are determined based on the target vector register size and pack size. 
For each candidate factor, we virtually unroll the loop and pack vectorizable instructions. The optimal unrolling factor is selected using a cost function (Section\ref{vir:cost-function}) that evaluates the benefits of vectorization after unrolling. 
Once loop unrolling is performed with the chosen factor, we rerun the packing pass and generate vector instructions from the {Instruction Packs}.

\subsubsection{Dependence Graph} \label{vir:dependence-graph}
\textit{Dependence graph} is designed to capture the dependence between \textit{Entries} in each layer's code list. Since \textit{Entry} in {\IRb} are categorized as \textit{Instruction Entry} and \textit{Loop Entry} which defines a layer, the dependence are represented at \textit{Instruction Entry} level and layer level.

\paragraph{\textbf{\textit{Instruction Entry} Level Dependence.}}
We use a tree structure to represent the dependence relationships of fine-grained instructions. At the \textit{Instruction Entry} level, nodes in the dependence graph are categorized into three types:
\begin{itemize}
    \item \textit{Entity Node}: Represents a constant entity or a pointer.
    \item \textit{Control Predicate Node}: Represents a \textit{Control Predicate} associated with a specific item.
    \item \textit{Instruction Node}: Represents an instruction. A specialized subset is the \textit{Memory Reference Instruction Node}, which specifically denotes memory reference instructions.
\end{itemize}
Each node in the dependence graph is associated with a list of successors, representing the nodes it directly depends on.
Successors of an \textit{Entity Node} or \textit{Control Predicate Node} may include an \textit{Entity Node} representing a used entity or an \textit{Instruction Node} defining a referenced entity.
For an \textit{Instruction Node}, successors include a \textit{Control Predicate Node} and \textit{Entity Nodes} or \textit{Instruction Nodes} associated with entities used in the instruction. Additionally, \textit{Memory Reference Instruction Nodes} have \textit{memory reference successors}, representing dependencies on other \textit{Memory Reference Instruction Nodes}.

\paragraph{\textbf{Layer Level Dependence.}}
Each \textit{Loop Entry} defines a new layer. Dependencies between \textit{Loop Entries} and \textit{Instruction Entries} within the layer are captured in a layer-specific dependence map. These maps simplify the analysis of both intra-layer and inter-layer dependencies.

The \textit{dependence graph} can be constructed with data dependence analysis on {\IRb}.  

\subsubsection{Vector Packing} \label{sec:vector-packing}

We utilize a bottom-up approach on the tree structures in the \textit{Instruction Entry} 
 level \textit{dependence graph} to detect \textit{Vector Packs} .
 
\textit{Vector Packs} are sets of tree nodes that can be grouped for vectorization.
Corresponding to the different types of nodes in the dependence graph at \textit{Instruction Entry} level, \textit{Vector Packs} in {\IRb} are defined as follows:
\begin{itemize}
    \item \textit{Entity Pack}: A group of entities, such as constants or contiguous memory addresses. Packs of contiguous memory addresses can often be replaced by a single memory address with an expanded memory reference width.
    \item \textit{Control Predicate Pack}: A group of control predicates. Identical predicates can be reduced to one, representing specific control flow, while distinct predicates can form a predicate mask for efficient execution of vector instructions.
    \item \textit{Instruction Pack}: A group of independent and isomorphic instructions that can be packed together. Normally, operand and result of these instructions can form a \textit{Entity Pack} and their execution condition form a \textit{Control Predicate Pack}.
\end{itemize}

To identify \textit{Vector Packs}, we first determine \textit{Roots}, which are categorized as \textit{SLP Roots} and \textit{LLP Roots}. \textit{SLP Roots} are tuples of \textit{Store} instructions referencing contiguous memory, formed by iteratively merging pairs of adjacent \textit{Store} instructions. \textit{LLP Roots} reference contiguous memory across iterations.

For each \textit{Root}, we perform an upward search along the successors of its associated \textit{Instruction Nodes}. 
The dependence trees of these \textit{Instruction Nodes} are traversed in parallel, starting from the first successors of each \textit{Instruction Node}, if these nodes match and can form a \textit{Vector Pack}, we further explore the sub-trees rooted at them and try packing nodes along the way. 
In Figure \ref{fig:vectorization}, the four trees in the dashed box root at \textit{Store Instruction Nodes} forming a \textit{SLP Roots} and all the nodes on these trees can be fully packed.

\subsubsection{Packing with Extensions} \label{sec:packing-extension}

In the packing process, extensive extensions can be applied to detect more {Vector Packs}. 

\paragraph{\textbf{Instruction Equivalence Transformation.}}

A key requirement for packing different instructions is isomorphism. \textit{Instruction Nodes} for heterogeneous instructions fail the matching test will not be packed. 
In Figure \ref{fig:vectorization}, for example, The constructed tree for \texttt{b[0] = a[0]}, is heterogeneous to the other three statements. Even though the \textit{Store Instruction Nodes} are detected as a \textit{SLP Roots}, the successors fail the matching due to the instruction used to produce the value for the store operation. Similar problems are also reported in LLVM Bugzilla as missed vectorization opportunities~\cite{llvm-bug-31572, llvm-bug-30787}.

However, by applying equivalence transformations, we can create more efficient vector instructions. 
As illustrated in Figure \ref{fig:vectorization}, by introducing an equivalent transformation, such as \texttt{a[0] \textless{}\textless{} 0}, we can full pack the four trees. The transformation \texttt{a[0] \textless{}\textless{} 0} is semantically equivalent to \texttt{a[0]} because shifting by zero leaves the value unchanged, thus preserving correctness while enhancing vectorization opportunities.

Similar equivalence transformations can also be applied to successors of \textit{Control Predicate Nodes} to generate masks for vector instructions.

\paragraph{\textbf{Inter-loop Pack Detection.}}

Our packing strategy allows for the packing of instructions from different loops.

\begin{figure}[htbp]
\centering
\includegraphics[width=1\linewidth]{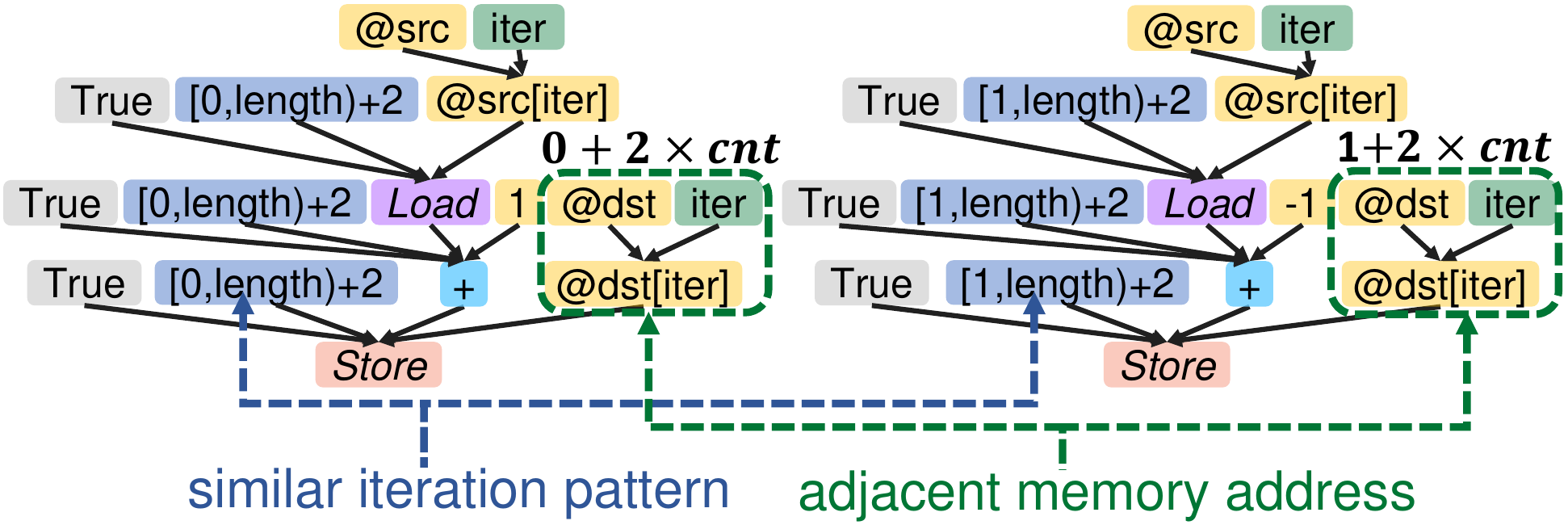}
\caption{Two complementary loops with interleaved access to the same array in Figure \ref{fig:motiv:fusion-unroll}. Inter-loop Pack Detection enables us to pack these two \textit{Store} instructions as an \textit{LLP Root}.}
\label{fig:inter-loop-pack}
\end{figure}

Figure \ref{fig:inter-loop-pack} illustrates an example to fuse, unroll and vectorize the two loops in Figure \ref{fig:motiv:fusion-unroll}. These two loops operate on the same arrays. Each loop accesses discontinuous memory addresses but performs similar operations on complementary array elements. The memory addresses accessed at each iteration for each loop are closely related to the loop iteration count, \(cnt\) in Figure \ref{fig:inter-loop-pack}. 
The \textit{Iterator} associated with each \textit{Entry} helps resolve these addresses and detect vectorization opportunities seamlessly.

For \textit{Store} instructions from different loops, we first determine if the \textit{Iterator} of their corresponding \textit{Entry} is resolvable. If so, we derive the memory addresses based on the iteration count. In the example from Figure \ref{fig:inter-loop-pack}, one loop accesses \(0 + 2 \times cnt\), while the other accesses \(1 + 2 \times cnt\) for the same arrays. This indicates that the \(cnt\)-th iterations of the loops target adjacent memory addresses, making their \textit{Store} instructions candidates for packing as a \textit{LLP Root}. Such fusion and vectorization can be achieved even if the loops are not adjacent, provided there are no dependencies between them, resulting in high-performance vectorized code.

\subsubsection{Cost Function} \label{vir:cost-function}
To decide whether we should transform certain \textit{Instruction Packs} into vector instructions, we evaluate the profitability of the packing using a cost function:
$$
Cost(p) = C_{vector}(p) - C_{scalar}(p) + penalty(p)
$$
In this formula, $C_{vector}$ represents the execution cost of the resulting vector instruction, $C_{scalar}$ is the aggregate cost of the scalar instructions before vectorization and $penalty$ accounts for the additional overhead introduced by data movement between vector and scalar registers.
We decide to transform the \textit{Instruction Pack} only if the cost is less than zero. 

Additionally, to choose optimal unrolling factors for loops, we estimate the cost of vectorization on the unrolled loops with the sum of cost of all the related \textit{Roots} we decide to vectorize.
Let $\mathcal{R}$ be all these \textit{Roots}, the total cost is estimated with
$$
\text{TotalCost} = \sum_{p \in \mathcal{R}} \left( C_{\text{vector}}(p) - C_{\text{scalar}}(p) + \text{penalty}(p) \right)
$$

\subsubsection{Generating Vector Instructions}
The conversion from scalar to vector code on {\IRb} focuses on identifying must-execute root instructions such as Store instructions. 

As shown in Algorithm \ref{alg:vec-function-transform}, we first construct a new function that is the vectorized version of the initial one. 
The new function begins with an empty body and meta data that includes the function parameters. 
Next, we vectorize the original function body to populate the new function body. 
This process starts with a sequential traversal of the flat code list at layer \texttt{0} which is outside any nested loop. 
In line \texttt{3} a placeholder is inserted into the \texttt{layerMap}, where the layer index is used as key and loop information as value, because layer \texttt{0} is not a loop. 
The loop from line \texttt{4} to line \texttt{15} then traverses the entries in the code list.

\definecolor{darkblue}{RGB}{0, 0, 153}
\definecolor{darkred}{RGB}{153, 0, 0}
\renewcommand{\KwSty}[1]{\textnormal{\textcolor{darkblue}{\bfseries #1}}\unskip}
\begin{algorithm}[t]
\scriptsize
\caption{Transformation to Vector Function}
\label{alg:vec-function-transform}
\KwData{$function$: input scalar function}
\KwResult{$vecFunction$: transformed vector function}
\tcp{\textcolor{darkred}{Initialize vector function and order manager}}
$vecFunction \leftarrow$ \textbf{new} VecFunction($function$)\;
$orderManager.init(vecFunction)$\;
\tcp{\textcolor{darkred}{Insert placeholder for code list}}
$layerMap.setValue([0], null)$\;
\tcp{\textcolor{darkred}{Transform code list}}
\For{$entry$ \textbf{in} $function.codeList$}{
    \If{$entry$ \textbf{is instance of} InstEntry}{
        $instruction \leftarrow entry.getInst()$\;
        \tcp{\textcolor{darkred}{Transform root instructions}}
        \If{$instruction$ \textbf{is instance of} StoreInst \textbf{or} FuncCallInst}{
            $transformToVector(instruction)$\;
            $orderManager.popStack()$\;
        }
    }
    \If{$entry$ \textbf{is instance of} LoopEntry}{  
        $loop \leftarrow entry.getLoop()$\;
        \tcp{\textcolor{darkred}{Transform loop body}}
        $transformCodeList(loop.loopCodeList)$\;
        \tcp{\textcolor{darkred}{Transform loop condition}}
        $manageDependency(loop.condition)$\;
        $orderManager.popStack()$\;
        $layerMap.getValue([loop.getId()]).condition = getControlPredicate(loop.condition)$\;
    }
}
\tcp{\textcolor{darkred}{Transform return value if it exists}}
\If{$function.retVal \neq null$}{
    \tcp{\textcolor{darkred}{Find the instruction defining return value}}
    $instruction \leftarrow function.getEntry(function.retVal.getDef()).getItem()$\;
    \tcp{\textcolor{darkred}{Transform the instruction}}
    $transformToVector(instruction)$\;
    $orderManager.popStack()$\;
    $vecFunction.retVal \leftarrow getEntity(function.retVal)$\;
}
\Return{$vecFunction$}\;
\end{algorithm}

If the \texttt{entry} is an \textit{Instruction Entry}, we checks whether its \textit{Instruction Item} qualifies as a root instruction. 
In such cases the instruction is transformed and inserted into the new function using the function \texttt{transformToVector}. 
This function decides whether the instruction should be packed and transformed into a vector instruction. 
Vectorizable instructions are transformed together with other instructions in the same pack. 
Before a transformed instruction is inserted into the code list, the correct insertion layer is determined and its dependencies are managed to ensure the proper execution order. 
The layer is determined by the \textit{Iterator} of the \textit{Instruction Entry}. If the \textit{Iterator} is \texttt{null}, the transformed instruction is inserted in layer \texttt{0}. Otherwise a nested loop corresponding to the \textit{Iterator} is either constructed or located and the transformed instruction is inserted into its body. The \texttt{layerMap} in Algorithm \ref{alg:vec-function-transform} stores the information of each layer.

Dependencies are handled recursively by creating or linking successors according to the corresponding tree or subtree in the \textit{dependence graph} introduced in Section \ref{vir:dependence-graph}. 
Creating a successor means transforming an instruction or a pack of instructions that have not been processed. 
Linking means finding the transformed instructions. 
The new instruction depends on these successors. Once all dependencies are resolved, the new instruction is placed after a ready point in the appropriate layer. 

A specialized data structure called the \textit{Order Manager} is used to manage instruction dependencies. This structure operates on a stack of dependency information. 
When a new \textit{Instruction Entry} is inserted, a \texttt{null} placeholder is pushed onto the stack. The \textit{Order Manager} then recursively resolves the dependencies. 
During recursion, the top of the stack is updated with the processed \textit{Instruction Entry} and the previous dependency information. After recursion completes the top of the stack is popped to determine the ready point for safely inserting the new \textit{Instruction Entry}. This process ensures that all dependencies are preserved.

If the \texttt{entry} is a \textit{Loop Entry}, the flat code list of the loop body is first transformed in the same manner as the loop from line \texttt{4} to line \texttt{15}. 
After that, the \textit{Instruction Entries} defining the loop exit condition are transformed. 
In line \texttt{13} of Algorithm \ref{alg:vec-function-transform}, the condition is recorded in the \texttt{layerMap} as the loop layer information.

Finally, if the function has a return value the instruction defining that value is transformed as well.

\subsection{Transforming {{\IRb}} to CFG-based IR}

\begin{figure}[htbp]
    \centering
    \includegraphics[width=1\linewidth]{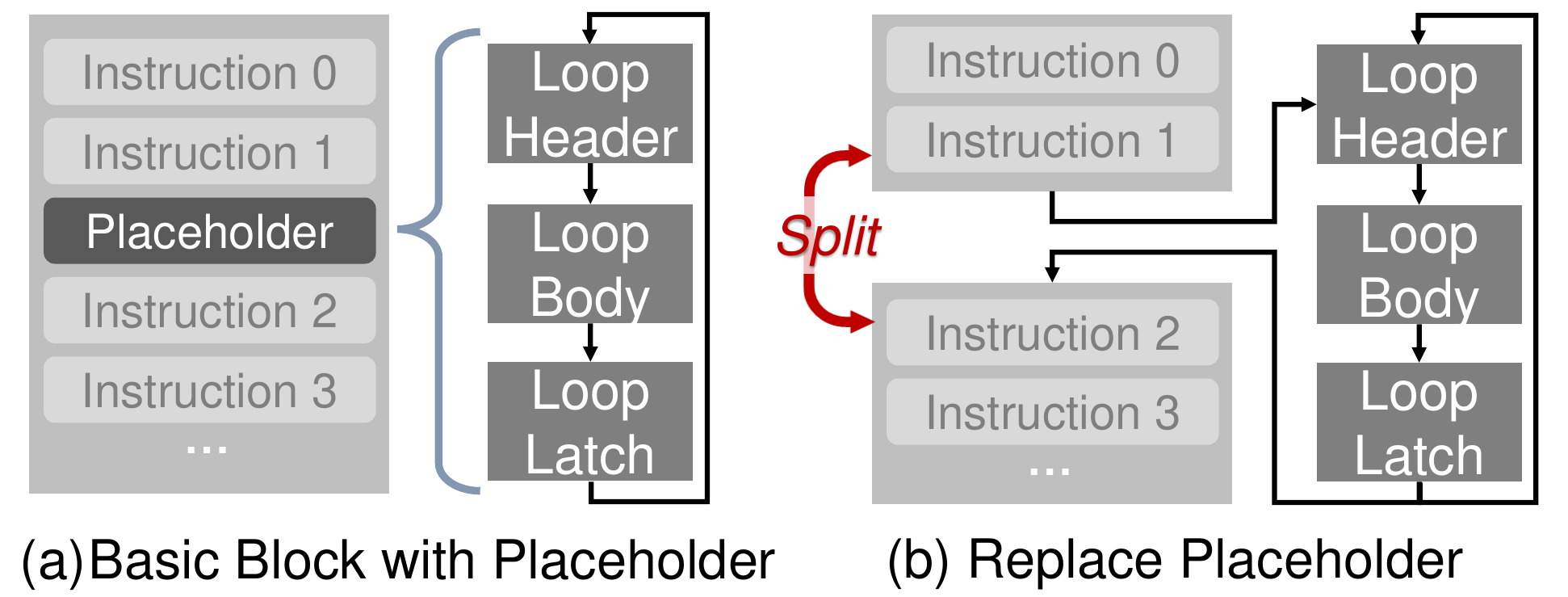}
    \caption{CFG Reconstruction Across Layers}
    \label{fig:cfg-reconstruction}
\end{figure}

The reconstruction of the control flow graph (CFG) in CFG-based IR, LLVM IR in our case, from {\IRb} focuses on building basic blocks. Using the dependence graph of the vectorized function, we apply a layer-wise control flow reconstruction approach. First, we construct branch control flows within each layer, inserting placeholders for loops to treat them as regular instructions. As shown in Figure \ref{fig:cfg-reconstruction}, once branch control flows for all layers are reconstructed, we connect the layers by splitting basic blocks at the placeholders. The first block ends with a \textit{Jump} to the loop header, while the \textit{Branch} in the loop latch block is updated to point to both the second block and the loop header.

The reconstruction process traverses the straight-line code list of each layer, initiating recursive transformations from "must-execute" root instructions, such as \textit{Store} and \textit{Function Call}, following the dependence tree.

Although this approach may initially complicate the control flow, subsequent optimization passes on LLVM IR efficiently simplify it, resulting in an optimized structure.

\section{Evaluation}

We evaluated the effectiveness of {\IRb}'s vectorization program and analyzed its sources of improvement.

\paragraph{Setup.}
We used Clang 17 and GCC 11 as baselines, compiling target programs to the x86 AVX2 instruction set with the options \texttt{-mavx2 -O3}. Additionally, inlining was disabled to measure the vectorizable kernel execution time more accurately. The testing platform was an Intel Ultra 7 PC with 96GB of memory. Turbo boost was disabled to ensure stable execution times. Test cases were selected from \texttt{tsvc} and included specially rewritten real-world image pixel processing code that existing compilers fail to vectorize.

\subsection{Vectorization Effectiveness}

\begin{figure}[htbp]
    \centering
    \includegraphics[width=\columnwidth]{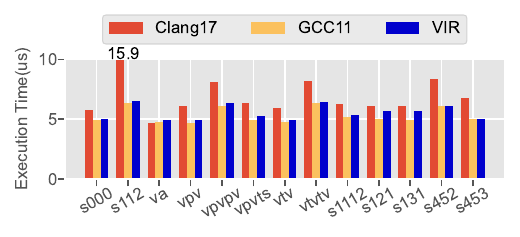}
    \caption{Execution time for {\IRb}, Clang, and GCC under \texttt{tsvc}.}
    \label{fig:eval:vectorization-effectiveness}
\end{figure}

The results for \texttt{tsvc} demonstrate that {\IRb} can further optimize LLVM's performance, as shown in \autoref{fig:eval:vectorization-effectiveness}. In the \texttt{tsvc} vectorization test cases, {\IRb} improves execution performance by 27\% compared to LLVM IR, with an average improvement of 15\%. Compared to GCC, performance decreases by only 5\% on average. Specifically, for the \texttt{s112} test case, LLVM cannot vectorize it correctly. In contrast, {\IRb} reduces the overhead by 60\%, achieving performance close to that of GCC.

\begin{figure}[htbp]
    \centering
    \includegraphics[width=\columnwidth]{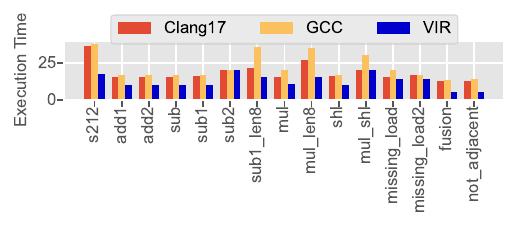}
    \caption{Execution time for {\IRb}, Clang, and GCC under specially rewritten real-world image pixel processing code.}
    \label{fig:eval:vectorization-special}
\end{figure}

In specially rewritten real-world image pixel processing code, which existing compilers fail to vectorize, {\IRb} achieves up to a 53\% improvement over LLVM, with an average improvement of 27\%, as shown in \autoref{fig:eval:vectorization-special}. Compared to GCC, {\IRb} achieves up to a 58\% improvement, with an average of 39\%. This performance gain in test cases like \texttt{sub} results from maintaining symmetry. Compiler optimizations, such as dead code elimination or constant propagation, often disrupt the symmetrical mathematical structure, which can harm vectorization. {\IRb} preserves this symmetry and successfully vectorizes the instructions. For test cases like \texttt{mul\_shl} and \texttt{fusion}, performance gains result from extensions explained in Section \ref{sec:packing-extension} on isomorphism detection. Specifically, equivalence transformation leads to high-performance vectorized code for \texttt{mul\_shl}, and inter-loop pack detection enhances performance for \texttt{fusion}.

\subsection{Case Study: Isomorphic Expansions}

\begin{figure}[htbp]
    \centering
    \includegraphics[width=\columnwidth]{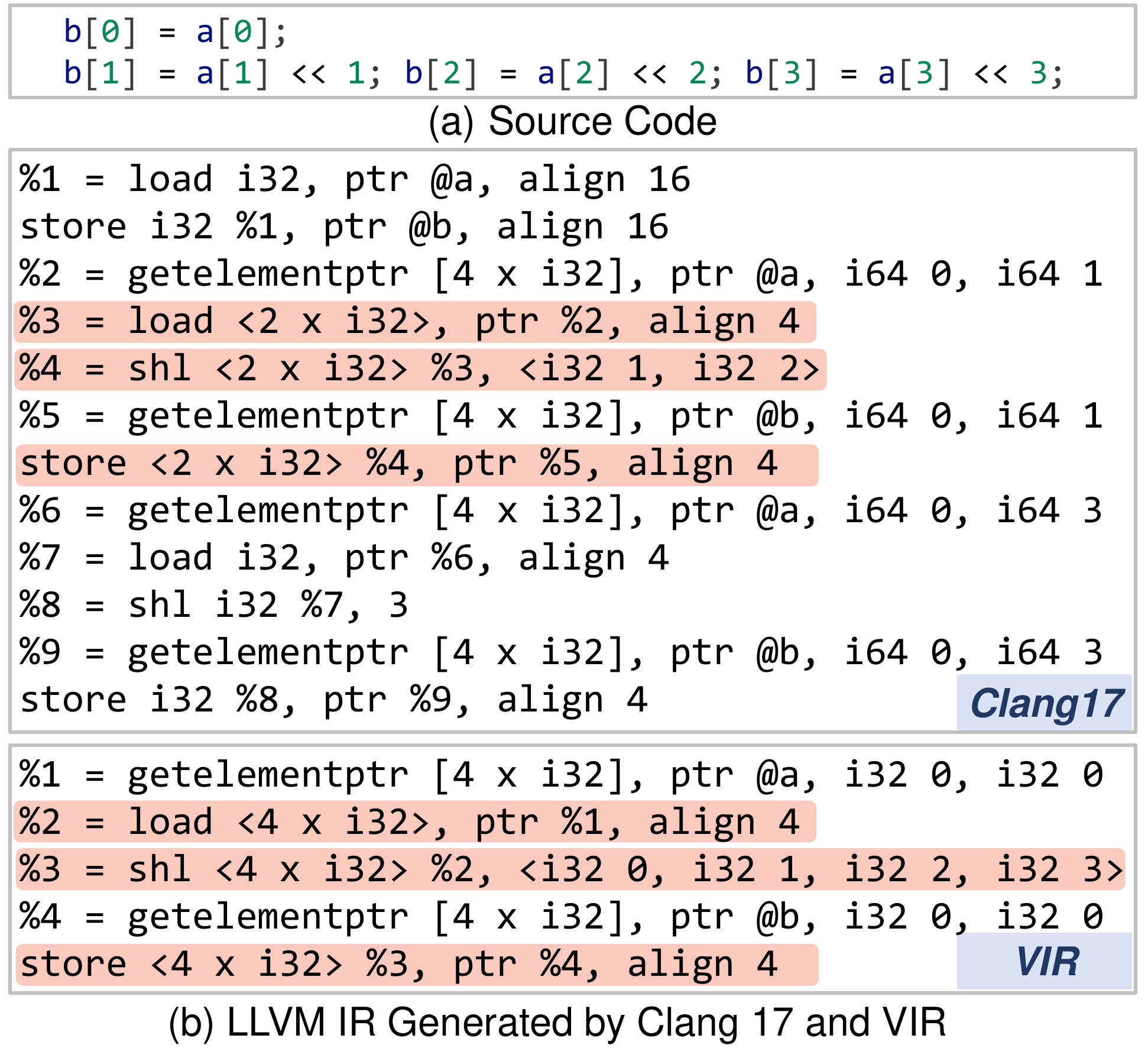}
    \caption{Case of equivalent transformations to perform isomorphic expansions.}
    \label{fig:eval:vir-llvm}
\end{figure}

To further explore the sources of {\IRb}'s performance improvement over LLVM, we analyzed one test case of isomorphic expansions, comparing the generated binary code with LLVM IR, as shown in \autoref{fig:eval:vir-llvm}. 

This structural uniformity introduced by {\IRb} significantly enhances the compiler's ability to perform vectorization, as it allows the compiler to recognize and exploit parallelism more effectively. 
It can expand the first statement (\texttt{b[0] = a[0];}) into statements with the same structure as subsequent ones, such as \texttt{b[i] = a[i] \textless{}\textless{} i;}, enabling a wider range of vectorization.
By aligning the first statement with the subsequent ones, {\IRb} removes irregularities and simplifies the loop structure, enabling LLVM's backend to leverage SIMD instructions more efficiently. 
Consequently, this optimization results in improved runtime performance, highlighting the importance of uniform code structure in facilitating advanced compiler optimizations such as vectorization.

\subsection{Case Study: Real-world Application}
{Common examples of programs exhibiting vectorizable characteristics include ray tracing and rendering in computer graphics (CG), 
as well as matrix computations in large language model~(LLM) inference. 
Due to the highly standardized and repetitive nature of operators used in LLMs, 
hardware vendors have already provided sufficiently optimized interfaces tailored specifically for model developers. 
In contrast, rendering operations in computer graphics often involve diverse and customized ray tracing procedures that require developers to define their own implementations. 
To investigate this further, we conducted tests to evaluate the rendering performance and effectiveness of raylib~\cite{raylib2025rtextures}.}

\begin{figure}[!htbp]
    \centering
    \includegraphics[width=\columnwidth]{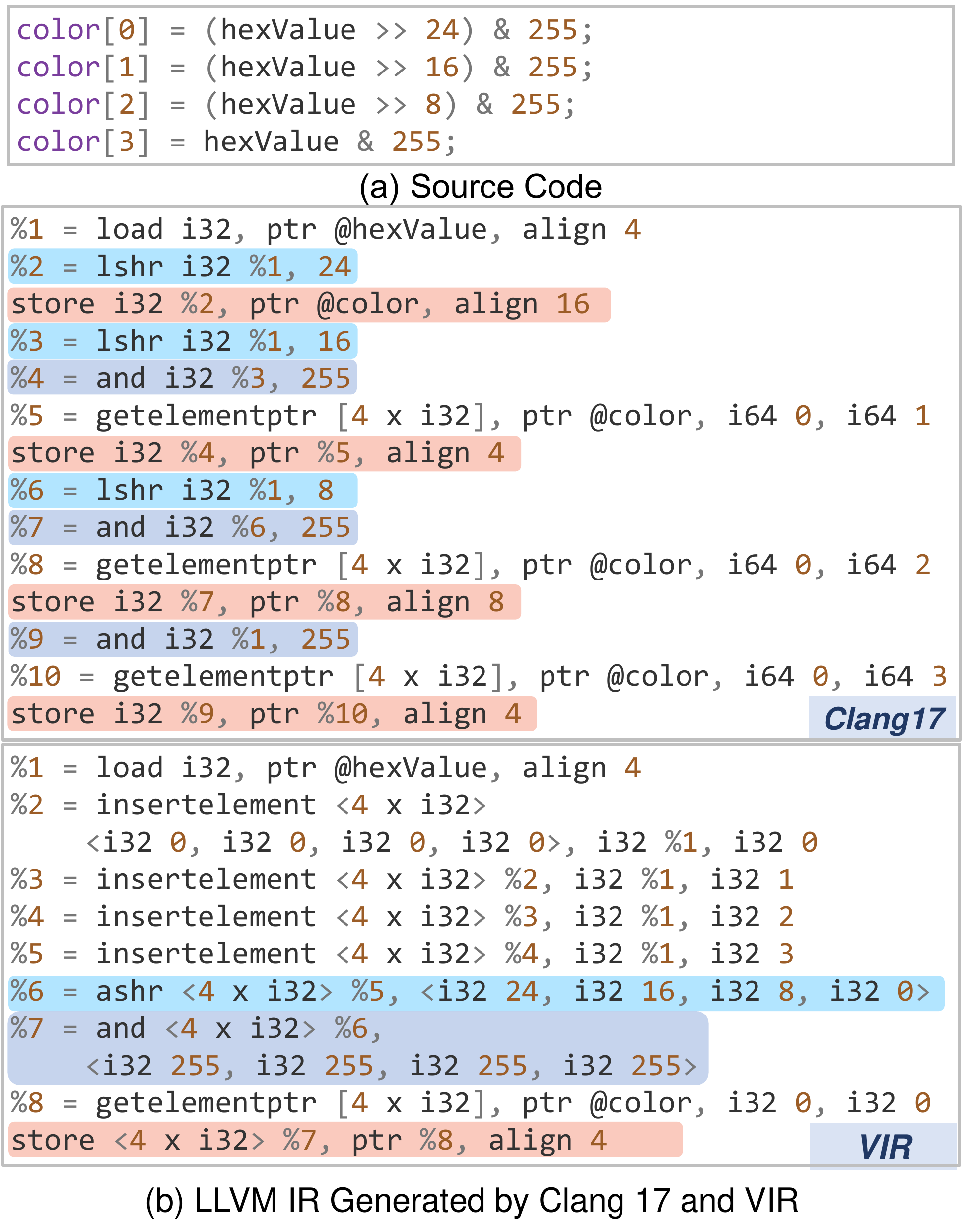}
    \caption{A real word case from raylib~\cite{raylib2025rtextures}.}
    \label{fig:eval:vir-llvm-real-world}
\end{figure}

{In the code snippet shown in~\autoref{fig:eval:vir-llvm-real-world}, 
the Clang compiler fails to vectorize instructions that involve combinations of bitwise operations. 
This occurs because Clang cannot accurately recognize the isomorphism relationships among these operations. 
Besides, it further perform some scalar optimizations including skipping \texttt{\& 255} for the first statement (\texttt{color[0] = (hexValue \textgreater{}\textgreater{} 24) \& 255;}), which further destroy the isomorphism.
Thus, even when vectorization flags are enabled, it is unable to vectorize such code, leading to sub-optimal performance. 
In our system, the isomorphism within this portion of the code have been fully reconstructed, enabling successful vectorization.}

\section{Discussion}
\paragraph{Integrating with Clang Toolchains.}
Our work can be integrated into the Clang toolchain by performing additional modifications and enhancements directly on Clang's Abstract Syntax Tree (AST)~\cite{clang-ast}. Specifically, we only need to further prune the AST to remove unnecessary nodes and annotate loop nodes with additional attributes that reflect our optimization or analysis requirements. These additional attributes can provide essential metadata to guide subsequent compilation stages, enabling more precise and targeted optimizations at later phases.
Moreover, the modifications we proposed at the LLVM IR can be encapsulated into a standalone IR pass. 
This IR pass can be seamlessly integrated into the existing LLVM compilation pipeline, automatically updating and refining LLVM IR instructions to reflect our optimization strategies. 

\paragraph{Language Generalization.}
Our approach does not impose restrictions on the type of input programming languages. Specifically, our method is compatible with all languages that can be compiled and translated into LLVM IR. This broad compatibility includes widely-used programming languages such as C and C++, as well as any other languages supported by LLVM's compilation infrastructure. As LLVM IR serves as a common intermediate representation for numerous languages, our approach leverages this flexibility, enabling developers to seamlessly integrate our techniques into their existing workflows without requiring significant modifications or specialized language-specific adjustments. 

\paragraph{Application Generalization.}
Our approach is neither limited nor dependent on simple shift operations or abbreviated syntactic sugar.
Instead, it remains applicable even to code structures that exhibit high complexity. By design, this method generalizes effectively, ensuring robustness and flexibility when handling sophisticated syntax and elaborate compiler or operating system constructs.

\section{Related Work}

\paragraph{Vectorization.}
Vectorizing programs can significantly enhance the utilization of SIMD components, thereby improving execution efficiency. 
Manualy vectorize program requires excessive effort by human experts, calling for automatic aproaches.
Automatic vectorization has been a longstanding focus of research. Allen and Kennedy \cite{DBLP:journals/toplas/AllenK87} established the foundations of loop vectorization, transforming loop iterations to execute simultaneously using SIMD instructions. 
Larsen and Amarasinghe \cite{larsen2000exploiting} introduced superword level parallelism (SLP), enabling vectorizing instrctions within a basic block. 
Subsequent works \cite{nuzman2008outer,DBLP:conf/pldi/NuzmanRZ06,DBLP:conf/pldi/BaghsorkhiVW16,chen2021vegen,karrenberg2015whole,nuzman2011vapor,barthe2013relational,porpodas2018look} have further advanced automatic vectorization. 
VALU \cite{rocha2020vectorization} introduced a vectorization-aware loop unrolling heuristic. 
SuperVectorization \cite{chen2022all} proposed a novel approach which simplifies code motion to vectorize instructions across basic blocks. 

\paragraph{Intermediate Representation.}
Intermediate Representation (IR) is a crucial abstraction in compilers, positioned between binary code and the abstract syntax tree. 
CFG-based IR, such as those used in LLVM ~\cite{DBLP:conf/cgo/LattnerA04} and GCC~\cite{gcc}, organize instructions in a graph structure where nodes represent basic blocks and edges denote control flow, facilitating conventional control-flow-sensitive analyses and optimizations.
To enable more advanced transformations, such as code motion and parallelization, it is necessary to analyze not only control dependencies but also data dependencies among instructions. Program Dependence Graphs (PDG) \cite{ferrante1987program} provide a representation of these dependencies, combining both control and data flow.
Building on these ideas, the Static Single Assignment (SSA) form was introduced by Cytron et al. \cite{cytron1991efficiently} as a practical representation that makes data dependencies explicit and simplifies various compiler optimizations. 
SSA is now widely adopted in modern compilers such as LLVM \cite{DBLP:conf/cgo/LattnerA04}, enabling optimizations like constant propagation \cite{wegman1991constant} and dominance-based analyses \cite{lengauer1979fast}.

Traditionally, vectorization is performed by optimization passes~\cite{llvm-vectorizers} on CFG-based IR like LLVM IR~\cite{DBLP:conf/cgo/LattnerA04}. 
However, CFG-based IR complicate code motion, which is essential for vectorization.
To address this challenge, Predicated SSA (PSSA)\cite{chen2021vegen, chen2022all} was developed. PSSA replace CFG with flat code list and makes code motion easier, enabling control-flow vectorization with SLP.

\section{Conclusion}
In this paper, we present a novel vectorization pipeline that addresses key limitations in existing compiler frameworks. Our approach specifically targets the problem of disjoint vectorization passes and the lack of extensibility, which often hinder optimization opportunities.
By proposing two specialized intermediate representations, {\IRa} for capturing high-level structural information and {\IRb} for explicitly encoding instruction dependencies, we significantly enhanced the compiler's ability to identify and exploit vectorization opportunities. 
Experimental evaluations demonstrate that our approach achieves substantial performance improvements, outperforming LLVM and GCC by up to 53\% and 58\%, respectively, highlighting the effectiveness and potential of our IR-based vectorization strategy for modern SIMD architectures.

\bibliographystyle{ACM-Reference-Format}
\bibliography{sample-base}

\appendix









\end{document}